\documentclass[iop, twocolumn]{aastex631}
\hypersetup{citecolor=blue,urlcolor=red}
\usepackage{multirow}
\usepackage{amsmath}
\usepackage{graphicx}
\usepackage{subfigure}
\usepackage{hyperref}  
\usepackage{nameref}
\usepackage{threeparttable} 
\shorttitle{PSR~B1257+12}
\shortauthors{Yao et al.}
\begin{document}
\title{Scintillation of the first-known pulsar planetary system}
\correspondingauthor{Jumei Yao, Lei Zhang, Di Li}
\email{yaojumei@xao.ac.cn, leizhang996@nao.cas.cn, dili@tsinghua.edu.cn}
\author{J. M. Yao\href{https://orcid.org/0000-0002-4997-045X}}
\affiliation{State Key Laboratory of Radio Astronomy and Technology, Xinjiang Astronomical Observatory, CAS, 150 Science 1-Street, Urumqi, Xinjiang, 830011, P. R. China}
\affiliation{Xinjiang Key Laboratory of Radio Astrophysics, 150 Science 1-Street, Urumqi 830011, P. R.  China}
\author{L. Zhang}
\affiliation{State Key Laboratory of Radio Astronomy and Technology, National Astronomical Observatories, CAS, Beijing 100101, China}
\affiliation{Centre for Astrophysics and Supercomputing, Swinburne University of Technology, Hawthorn, VIC 3122, Australia}
\author{A. Wolszczan}
\affiliation{Department of Astronomy and Astrophysics, Pennsylvania State University, 525 Davey Laboratory, University Park, PA 16802, USA}
\affiliation{Center for Exoplanets and Habitable Worlds, Pennsylvania State University, 525 Davey Laboratory, University Park, PA 16802, USA}
\author{William A. Coles}
\affiliation{Electrical and Computer Engineering, University of California, San Diego, 92093, USA}
\author{D. Li}
\affiliation{New Cornerstone Science Laboratory, Department of Astronomy, Tsinghua University, Beijing 100084, China}
\affiliation{State Key Laboratory of Radio Astronomy and Technology, National Astronomical Observatories, CAS, Beijing 100101, China}
\author{Richard N. Manchester}
\affiliation{Australia Telescope National Facility, CSIRO Space and Astronomy, P.O. Box 76, Epping NSW 1710, Australia}
\author{N. Wang}
\affiliation{State Key Laboratory of Radio Astronomy and Technology, Xinjiang Astronomical Observatory, CAS, 150 Science 1-Street, Urumqi, Xinjiang, 830011, P. R. China}
\affiliation{Xinjiang Key Laboratory of Radio Astrophysics, 150 Science 1-Street, Urumqi 830011, P. R.  China}
\author{C. H. Niu}
\affiliation{Institute of Astrophysics, Central China Normal University, Wuhan 430079, China}
\author{P. Wang}
\affiliation{State Key Laboratory of Radio Astronomy and Technology, National Astronomical Observatories, CAS, Beijing 100101, China}
\author{F. F. Kou}
\affiliation{State Key Laboratory of Radio Astronomy and Technology, Xinjiang Astronomical Observatory, CAS, 150 Science 1-Street, Urumqi, Xinjiang, 830011, P. R. China}
\affiliation{Xinjiang Key Laboratory of Radio Astrophysics, 150 Science 1-Street, Urumqi 830011, P. R.  China}
\author{J. P. Yuan}
\affiliation{State Key Laboratory of Radio Astronomy and Technology, Xinjiang Astronomical Observatory, CAS, 150 Science 1-Street, Urumqi, Xinjiang, 830011, P. R. China}
\affiliation{Xinjiang Key Laboratory of Radio Astrophysics, 150 Science 1-Street, Urumqi 830011, P. R.  China}
\begin{abstract}
We present a scintillation study of the first-known pulsar planetary system, PSR~B1257+12, using the Five-hundred-meter Aperture Spherical radio Telescope (FAST). A total of 31 observations with durations greater than or equal to 30 minutes were analyzed. For 14 longer observations (greater than or equal to 120 minutes), one-dimensional autocorrelation function analyses yielded the scintillation timescale, scintillation bandwidth, and frequency-drift rate for 12 epochs. Two observations show strong periodic modulation in the frequency-domain auto-correlation function, likely caused by astronomical-unit-scale structures along the propagation path, preventing reliable measurements of the scintillation timescale and bandwidth. In three observations, secondary spectra reveal simultaneous detections of inner, middle, and outer arcs. Analysis of the annual modulation of the inner-arc curvature indicates isotropic scattering, with a screen distance of $233\pm28$~pc and transverse velocity $V_{\rm scr,\alpha}=-7.16\pm2.16$ km~s$^{-1}$, $V_{\rm scr,\delta}=-41.07\pm5.69$ km~s$^{-1}$. Delay-profile analysis for both the inner and outer arcs suggest spectral exponents consistent with, or smaller than, the Kolmogorov value. Under isotropic scattering, the screen--pulsar distances are $354\pm22$~pc and $166\pm12$~pc for the middle and outer arcs. Combining the results from long-term timing analyses with our scintillation measurements, we find that the dispersion measure (DM) variations are primarily dominated by plasma located further away from the pulsar. The low DM-change rate of the outer arc and the absence of nearby scattering screens suggest that the immediate environment of the pulsar may be relatively clean. Alternatively, scattering screens closer to the pulsar may exist but remain undetected, requiring higher-sensitivity or longer-duration observations.
\end{abstract}
 
\section{Introduction}\label{sec:Intro}
The discovery of planetary systems around pulsars represents a milestone in modern astrophysics, as it demonstrates that planets can survive or even form in the exotic environments. Since the discovery of the first pulsar planetary system, PSR~B1257+12 \citep{1992Natur.355..145W}, which was also the first known extrasolar planetary system, several recent studies have confirmed the existence of six pulsar planetary systems containing a total of eight planets. All of these confirmed systems are millisecond pulsars (MSPs), as summarized by \cite{2025ApJ...982...63L}. Among them, only PSR~B1257+12 hosts multiple planets (3). At first glance, the presence of planets around a neutron star seems surprising; however, subsequent studies have demonstrated that the formation of such systems is not impossible to explain. According to \cite{2025ApJ...982...63L}, since planetary systems are unlikely to survive the supernova explosion that creates a neutron star, three alternative formation mechanisms have been proposed: (I) the planets form during the supernova, or soon afterward, from ejecta captured by the nascent neutron star; (II) planetary-mass bodies are captured later on; (III) stellar companions are so heavily ablated by the NS winds and radiation that their masses are reduced to planet scales. For PSR~B1257+12, since it hosts three planets and the outer two planets satisfy the near 3:2 mean motion resonance (MMR) \citep{2003ApJ...591L.147K}, the formation mechanism is unlikely to be (II) or (III). It's likely to have originated from (I), i.e., the circumpulsar gas disk formed from the material of the supernova explosion or the later fallback material, which then formed the planets. As proposed by \cite{2007ApJ...666.1232C} and \cite{2009ApJ...691..382H}, the circumpulsar gas disk could not only form from fallback material, but it may also form from interactions with a binary companion due to tidal disruption. If these planets originated from such a disk, then there may be residual material surrounding the planetary system. Therefore, we aim to investigate the ionized medium around PSR~B1257+12 and along its signal propagation path based on observations from FAST \citep{2011IJMPD..20..989N, 2018IMMag..19..112L}.

PSR~B1257+12, with a dispersion measure of 10.1$\pm$0.2~pc~cm$^{-3}$, was discovered in a high-galactic-latitude survey for millisecond pulsars conducted with the Arecibo telescope \citep{1990IAUC.5073....1W}. Based on follow-up timing observations with Arecibo, \cite{1992Natur.355..145W} first discovered the two outer planets, and later, \cite{1994Sci...264..538W} identified the inner planet. According to \cite{2003ApJ...591L.147K}, the outer two planets, designated C and D, have semimajor axes of 0.36 and 0.46~AU, masses of 4.3 and 3.9~$M_\oplus$, and orbital periods of 66.5 and 98.2~days, respectively. The inner planet, designated B, has a semimajor axis of 0.19~AU, a mass of 0.02~$M_\oplus$, and an orbital period of 25.3~days. From \cite{2003ApJ...591L.147K}, the proper motion of PSR~B1257+12, derived from timing analysis, is 45.50$\pm$0.04~mas~yr$^{-1}$ in right ascension and $-$84.70$\pm$0.07~mas~yr$^{-1}$ in declination. Based on observations with Very Long Baseline Array and the European Very Long Baseline Interferometry (VLBI) Network at 1465 and 1540~MHz, for PSR~B1257+12, \cite{2013MNRAS.433..162Y} measured the proper motion of $\mu_{\alpha}=46.44\pm0.08$~mas~yr$^{-1}$ and $\mu_{\delta}=-84.87\pm0.32$~mas~yr$^{-1}$, and the parallax distance of D$=$710$^{+43}_{-38}$~pc. From \cite{2019MNRAS.484.3646S}, the rotation measure (RM) of PSR~B1257+12, determined from Low-Frequency Array observations, is 7.78$\pm$0.07~rad~m$^{-2}$. 

Pulsar scintillation studies serve as an important tool for probing the pulsar’s circumstellar environment, the properties of the ionized medium along the propagation path, and the intrinsic characteristics of the pulsar itself. Before the discovery of scintillation arcs, the autocorrelation function (ACF) of the dynamic spectrum was commonly used to estimate scintillation parameters, including the scintillation timescale ($\Delta t_d$), scintillation bandwidth ($\Delta \nu_d$), and the frequency drift rate (dt/d$\nu$) \citep{1999ApJ...514..272B}. Since the first detections of scintillation arcs from the secondary spectra of four pulsars \citep{2001ApJ...549L..97S}, more than one hundred pulsars have been observed to exhibit scintillation arcs in recent years \citep{2022ApJ...941...34S, 2023MNRAS.518.1086M}. In recent years, high signal-to-noise ratio observations have not only enabled the detection of scintillation arcs from more pulsars but also made it possible to detect scintillation arcs originating from scattering screens located several parsecs (pc) or astronomical units (AU) away from the pulsars. With FAST observations, the detection of scintillation arcs for PSRs~J0538+2817 and B0656+14 has opened up new avenues for studying the turbulent properties of their associated supernova remnant (SNR) shells, located tens of parsecs away from the pulsars, and for constraining the pulsar's radial velocity based on the geometry of the supernova remnant, especially for PSR~J0538+2817 \citep{2021NatAs...5..788Y, 2022ApJ...939...75Y}. Among pulsars observed with FAST, PSR~B0355+54 has the closest detected scattering screen, located only~10$^5$~AU from the pulsar \citep{2024MNRAS.527.7568O}. This proximity closely aligns with the expected location of the pulsar’s bow shock. However with MeerKAT observations, \cite{2025NatAs...9.1053R} discovered the largest number of scintillation arcs detected from a single pulsar, 25 arcs, for PSR~J0437$-$4715. Among these, the outer four arcs correspond to scattering screens located much closer (5000~AU) to the pulsar. Based on the measured radial distance and velocity of the main shock inferred from the outermost arc in the secondary spectrum, \cite{2025NatAs...9.1053R} determined the shock geometry and the three-dimensional space velocity of the pulsar. Therefore, the detection of scintillation arcs originating from scattering screens close to the pulsar, along with the measurement of their curvature, helps reveal the pulsar’s circumstellar environment and constrain its velocity. In addition to the arc curvature, \cite{2023MNRAS.521.6392R} noted that the measurement of the skewness on the ACF or the displacement of the arc's apex from the origin can be used to derive the phase gradient across the entire scattering disk. Based on the phase gradient, we can then calculate the change rate of the DM resulting from the pulsar’s transverse motion. Additionally, for the detection of scintillation arcs originating from scattering screens close to the Earth, the annual modulation of the arc curvature due to the Earth's motion allows for the measurement of the precise location of the scattering screen, as well as the anisotropy of the scattering \citep{2025arXiv250419531H, 2025NatAs...9.1053R}. 

For PSR~B1257+12, the scintillation timescale and scintillation bandwidth measured from dynamic spectra at 327 and 430~MHz, based on observations with the Ooty Radio Telescope and the Arecibo Radio Telescope, were used to estimate the scintillation velocity \citep{2000ApJ...531..345G, 2002ApJ...581..495B}. They found a scintillation velocity of about 200~km~s$^{-1}$, which is relatively high for a millisecond pulsar. However, to date, no scintillation arc has been detected for this source. In this paper, we use high-sensitivity observations made with FAST at frequencies around 1250~MHz to study the circumstellar environment of PSR~B1257+12 planetry system and the properties of the scattering screens along the propagation path. The secondary spectra for PSR~B1257+12 show three distinct arcs. We use the curvature of these arcs to demonstrate that the scattering of the signal occurs in the shell of the Local Bubble for the inner arc, while for the middle and outer arcs, the scattering takes place at distances of 354$\pm$22~pc and 166$\pm$12~pc from the pulsar, respectively. For the outer arc, the rate of change of the DM is only 18\% of that for the inner arc. Additionally, for both the inner and outer arcs, in most of the observed epochs, the turbulent spectral index is flatter than the classic Kolmogorov spectrum.

The arrangement of our paper is as follows: In Section~\ref{sec:Obs}, we describe the FAST observations and data processing procedures. Section~\ref{sec:DS_ACF} presents the dynamic spectra and ACF results. In Section~\ref{sec:SS}, we provide the secondary spectra and the associated analysis. Section~\ref{sec:SL_DM} discusses the scattering screens and the DM variation. Finally, in Section~\ref{sec:SD}, we summarize our results and present our conclusions

\section{Observations and data processing}\label{sec:Obs}
From the FAST archive, between August and December 2022, dense observations of the pulsar PSR~B1257+12 were conducted to improve timing precision in the search for lower-mass planets. Among these data, observations at MJDs~59809 and 59839 revealed three distinct arc structures in the secondary spectrum: inner, middle, and outer arcs.
To probe scattering screens closer to the pulsar and to ensure good annual coverage for studying the inner arc, we carried out a series of longer-duration observations with FAST from October 2024 to August 2025. In total, 11 observations were obtained, nearly all with durations of at least 120 minutes, and the longest lasting up to 5 hours, the maximum continuous tracking time achievable by FAST for PSR~B1257+12. Considering the relatively long scintillation timescale of PSR~B1257+12 (as shown in Table~\ref{tab:obs_sp}), we restrict our ACF analysis to observations with durations no shorter than 120 minutes. For observations longer than 30 minutes, we list the UTC, MJD, and observation duration of those exhibiting arc structures in the secondary spectrum in the first three columns of Table~\ref{tab:inner}. In total, three observations (MJDs~59809, 59839, and 60597) show all three arc layers. Owing to its stronger signal, the inner arc is detected in all 31 observations listed in Table~\ref{tab:inner}.

For all 31 observations, the central beam of the 19-beam receiver of the FAST telescope was used, covering a frequency range from 1000 to 1500~MHz. Due to reduced sensitivity at the edges of the band, the effective bandwidth is 400~MHz, spanning from 1050 to 1450~MHz. During the following data processing, we use the analysis program {\sc dspsr} \citep{2011PASA...28....1V}\footnote{\url{http://dspsr.sourceforge.net}} and the {\sc psrchive} software package \citep{2012AR&T....9..237V}\footnote{\url{http://psrchive.sourceforge.net}} to process the data. The data for each channel, with a bandwidth of 0.488~MHz, were folded at the topocentric pulse period using a sub-integration time of 30~s, and the noise files were also folded accordingly. 
Then, we performed interference removal on the folded data for both the source and the noise, and conducted polarization calibration for all observations except those on MJDs~59809, 59839, and 59863 (for which no noise files were available). Subsequently, we folded the polarization and used {\sc psrflux} to obtain the data for generating the dynamic spectrum.

Following \cite{2020ApJ...904..104R}, to generate the secondary spectrum, we first performed equi-wavelength sampling on the dynamic spectrum. As described in Section~\ref{sec:DS_ACF}, since the observation duration of most PSR~B1257+12 observations (30~min) is close to its scintillation timescale, we did not apply a Hamming window to the equi-wavelength sampled data. Next, we performed a two-dimensional (2D) autocorrelation, followed by pre-whitening of the data using first-differences. Afterward, we constructed the secondary spectrum by applying a 2D Fourier transform, then took its squared magnitude. Finally, we applied an appropriate post-darkening filter to recover the best estimate of the secondary spectrum. In the secondary spectrum, the Nyquist frequencies are $f_t(\rm Nyquist)=16.7$~mHz, and $f_\lambda(\rm Nyquist)=5123.5$~m$^{-1}$

\section{Dynamic Spectra and Autocovariance Functions}\label{sec:DS_ACF}
\begin{figure}[!h]
\center
 \includegraphics[width=6.0 cm, angle=270]{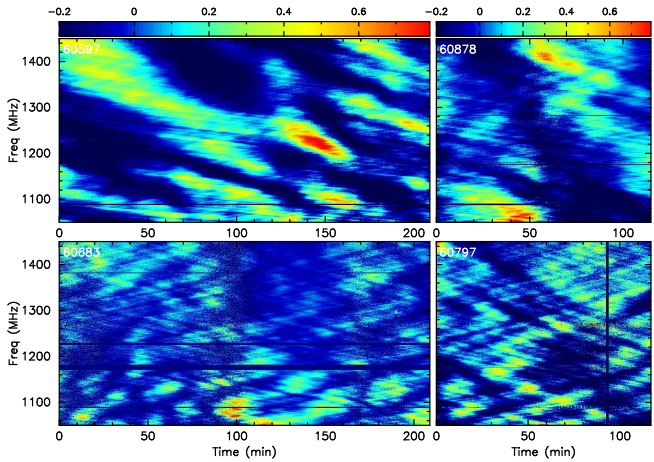}
    \caption{Dynamic spectra of PSR~B1257+12 observed with FAST over a 400~MHz bandwidth centered at 1250~MHz. The left panels show 210~min observations obtained on MJDs~60597 and 60683, while the right panels display 120~min observations from MJDs~60878 and 60797. All panels use identical scales for both axes, and the color scale is linear in signal power, shown in arbitrary units.}
\label{fig:DS_four}
\end{figure}

The dynamic spectra of PSR~B1257+12 exhibit two modes: a normal tilted patch pattern and a periodic fringe pattern. As shown in Figure~\ref{fig:DS_four}, the upper two panels display the normal tilted patch pattern for observations made on MJDs~60597 and 60878, while the lower two panels show the periodic fringe pattern for observations made on MJDs~60683 and 60797. As mentioned in \cite{yky+26}, these periodic fringe pattern arise from interference between each discrete sub-image and the rest of the image. The scintillation bandwidth and timescale could not be determined through ACF fitting. However, if the pulsar’s scattered image is dominated by two primary sub-images, a clear periodic modulation can appear in the ACF, particularly in the frequency-domain 1D ACF. For PSR~J1740+1000, \cite{yky+26} found that its multiple scattered images are produced by refraction from a discrete, dense AU-scale structure within the scattering region, which is very likely associated with the pulsar’s PWN. Therefore, for PSR~B1257+12, the pulsar's scattered multiple images may also originate from refraction by a discrete, dense AU-scale structure.

\begin{figure*}[!h]
\center
 \includegraphics[width=18.0 cm, angle=0]{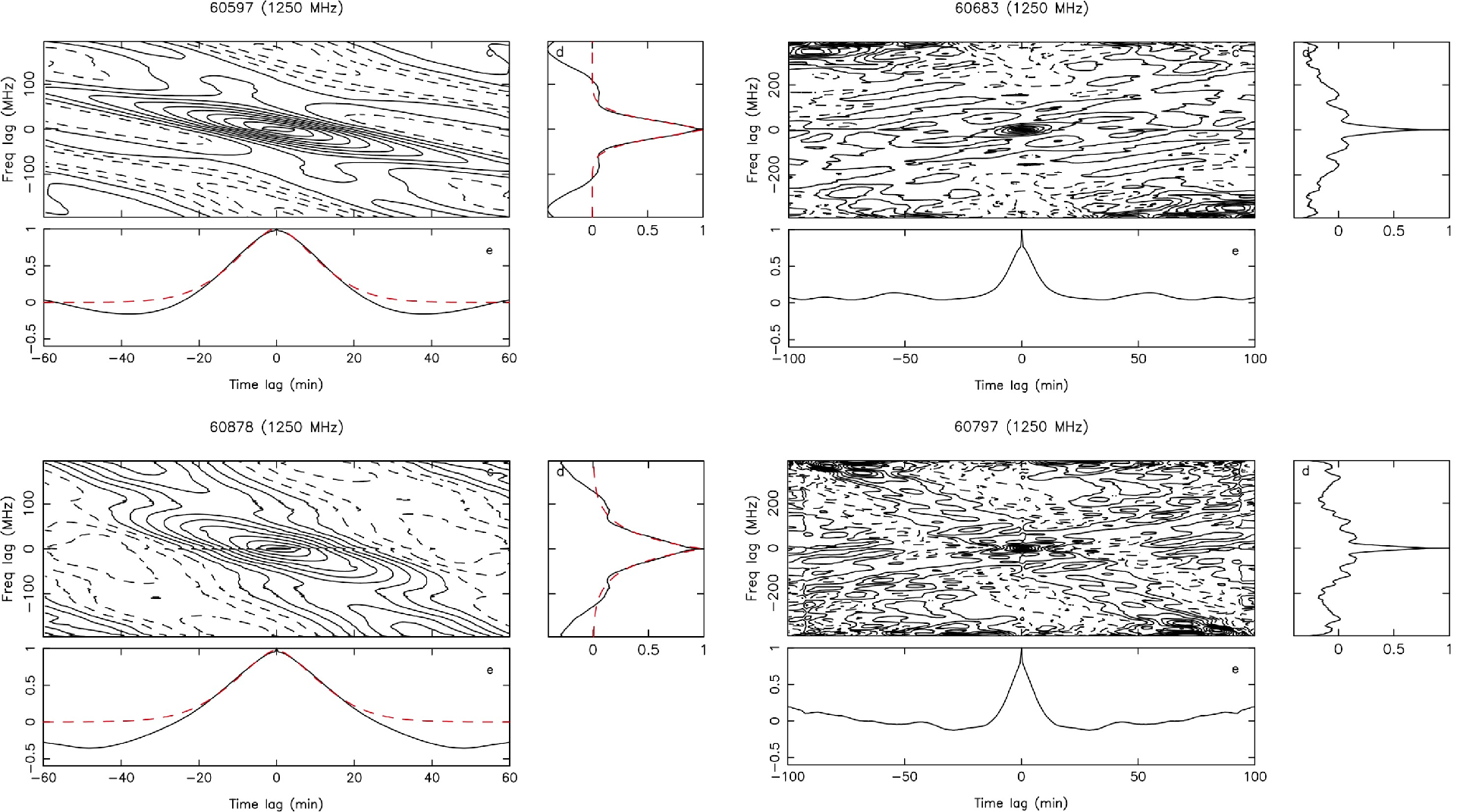}
    \caption{The ACFs for MJDs~60597, 60683, 60878, and 60797, along with the best-fit results for $\Delta \nu_d$ and $\Delta t_d$ at 1250~MHz for MJDs~60597 and 60878. Panel (c) shows the 2D ACF, panel (d) the 1D frequency-domain ACF, and panel (e) the 1D time-domain ACF. In panels (d) and (e) for MJDs~60597 and 60878, the red lines indicate the best-fit results for $\Delta \nu_d$ and $\Delta t_d$.}
\label{fig:acf_four}
\end{figure*}

\begin{table*}
\caption{Value and uncertainty of $\Delta t_d$, $\Delta \nu_{\rm d}$ and dt/d$\nu$ for PSR~B1257+12 at 1250~MHz.}
\centering
\setlength\tabcolsep{4.0pt}
\footnotesize
\begin{tabular}{ccccccc}
   \hline
      UTC& MJD & Time& A & $\Delta t_{\rm d} $& $\Delta \nu_{\rm d}$ & dt/d$\nu$ \\
   \hline
        &  &min &    & min & MHz& min~MHz$^{-1}$\\
   \hline
2022/08/18& 59809 & 180 & 0.870$\pm$0.006& 18.05$\pm$2.09&21.5$\pm$2.8&0.0\\
2022/09/17& 59839 & 180 & 0.631$\pm$0.019& 15.00$\pm$2.43&32.1$\pm$4.3&0.0 \\
2022/10/11& 59863 & 180 & 0.794$\pm$0.009& 21.2$\pm$3.53&39.3$\pm$6.6&0.0 \\
2024/10/14& 60597 & 210 & 1.014$\pm$0.007& 15.71$\pm$2.16&43.4$\pm$8.1&$-$0.440$\pm$0.012\\
2024/11/23& 60637 & 180 & 0.788$\pm$0.010& 18.00$\pm$2.45&28.5$\pm$4.5&$-$0.338$\pm$0.019\\
2025/01/08& 60683 & 210 & & & & \\
2025/02/27& 60733 & 210 & 0.973$\pm$0.002& 19.06$\pm$3.22&61.2$\pm$12.9&$-$0.418$\pm$0.008\\
2025/03/28& 60762 & 120 & 0.909$\pm$0.007& 12.47$\pm$1.18&15.4$\pm$2.3&$-$0.670$\pm$0.031\\
2025/04/23& 60788 & 120 & 0.812$\pm$0.015& 15.25$\pm$2.74&38.1$\pm$6.9&$-$0.090$\pm$0.025\\
2025/05/02& 60797 & 120 &  &  &  & \\
2025/05/14& 60809 & 120 & 0.780$\pm$0.016& 14.88$\pm$2.65&34.8$\pm$7.5&$-$0.429$\pm$0.014\\
2025/07/17& 60873 & 180 & 0.869$\pm$0.008& 14.83$\pm$1.37&14.8$\pm$1.8&$-$0.502$\pm$0.031\\
2025/07/22& 60878 & 120 & 0.987$\pm$0.005& 16.96$\pm$2.57&29.0$\pm$4.8&0.0\\
2025/08/17& 60904 & 290 & 0.795$\pm$0.013&16.61$\pm$1.60&19.6$\pm$2.5&0.0\\ 
   \hline
   \end{tabular}
   \label{tab:obs_sp}
\end{table*}

As mentioned in Section~\ref{sec:Obs}, for PSR~B1257+12, since the duration of most of the FAST observations is comparable to the scintillation timescale, we only performed ACF analysis on observations with a duration of at least 120~min. The UTC, MJDs and observation durations are listed in columns 1, 2 and 3 of Table~\ref{tab:obs_sp}. Among these observations, only two show the periodic fringe patterns, and we did not perform ACF fitting on these observations. For the remaining observations, following \cite{2025SCPMA..6839512W}, we first fit the 1D time-domain ACF using the following equation to obtain the parameters A and $\Delta t_d$:
\begin{equation}
\rho(0, \Delta t)=Ae^{-\left|\frac{\Delta t}{\Delta t_d}\right|^{\alpha_0}}(1-\frac{\left|\Delta t\right|}{T})
\label{eq:1D_acf_time}
\end{equation}
where $\rho(0, 0)=A+w=1$ (with A and w being parameters used to smooth the white noise peak at (0, 0)), and the exponent $\alpha_0=5/3$ corresponds to Kolmogorov turbulence. The triangle function $1-\left|\Delta t\right|/T$ is used to mitigate the influence of the finite observation length T. Next, we fix A and use the following equation to fit the 1D frequency-domain ACF, from which we obtain $\Delta \nu_d$ and $dt/d\nu$: 
\begin{equation}
\rho(\Delta \nu, 0)=Ae^{-\left(\left|\frac{0-\frac{dt}{d\nu}\Delta\nu}{\Delta t_d}\right|^{\frac{3\alpha_0}{2}}+\left|\frac{\Delta \nu}{\Delta \nu_d/ln2}\right|^{\frac{3}{2}}\right)^{\frac{2}{3}}}(1-\frac{\left|\Delta \nu\right|}{F})
\label{eq:1D_acf_freq}
\end{equation}
where the triangle function $1-\left|\Delta \nu\right|/F$ reduces the influence of finite observation bandwidth F. Since the 1D frequency-domain ACF fitting alone cannot constrain the sign of $dt/d\nu$, we combine Equations~\ref{eq:1D_acf_time} and \ref{eq:1D_acf_freq} and perform a 2D fit to the ACF peak region. The derived value of $dt/d\nu$ is subsequently used as the initial parameter for the 1D frequency-domain ACF fitting. For the dynamic spectra displayed in Figure~\ref{fig:DS_four}, the corresponding ACFs are shown in Figure~\ref{fig:acf_four}. In the left two panels, the red lines indicate the best-fit results. The resulting value and 1-$\sigma$ uncertainty of A, $\Delta t_d$, $\Delta \nu_d$ and $dt/d\nu$ are listed in columns 4, 5, 6 and 7 of Table~\ref{tab:obs_sp}. The uncertainty of $\Delta t_d$, $\Delta \nu_d$ includes two components: the statistical uncertainty from the data fitting and the fractional uncertainty due to the finite number of observed scintles in the dynamic spectra \citep{1985ApJ...288..221C, 2005MNRAS.358..270W}. From Table~\ref{tab:obs_sp}, the scintillation timescale ranges from 12 to 21~min, while the scintillation bandwidth fluctuates significantly, ranging from 15 to 60~MHz. This is understandable. According to \cite{2005MNRAS.358..270W}, for a Kolmogorov spectrum, the scintillation timescale evolves with a frequency index of 1.2, while the scintillation bandwidth evolves with a frequency index of 4.4. Due to the relatively small observation bandwidth and observation time, the scintillation timescale and scintillation bandwidth at the high frequency band (1450~MHz) are 1.5 and 4.1 times those at the low frequency band (1050~MHz), which exactly matches our measurements. For PSR~B1257+12, the mean values of $\Delta t_d$ and $\Delta \nu_d$ at 1250~MHz are 16$\pm$2~min and 31$\pm$13~MHz, respectively. Based on the mean value of scintillation bandwidth, we found that the scintillation strength, defined as $u=\sqrt{\nu/\Delta \nu_d}$ \citep{1990ARA&A..28..561R}, is 6.3. This indicates that, at 1250~MHz PSR~B1257+12 is in the intermediate regime, out of the weak scintillation regime, but not yet in the asymptotically strong regime. 

From the right panels of Figure~\ref{fig:acf_four}, on MJDs~60683 and 60797, the 1D frequency-domain ACF indeed shows strong periodic modulation, which is more pronounced than the periodic modulation in the 1D time-domain ACF. As calculated by \cite{yky+26} (see Equations~15 and 16), the ratio between the scintillation timescale and the modulation period in time is proportional to $\theta_g/\theta_d$ (where $\theta_g>\theta_d$, with $\theta_g$ being the refractive angle and $\theta_d$ the width of the angular spectrum). Meanwhile, the ratio between scintillation bandwidth and the modulation period in frequency is proportional to $\theta^2_g/\theta^2_d$, which makes the periodic modulation in frequency more prominent. 

\section{Secondary spectrum}\label{sec:SS}
\begin{figure}[!h]
\center
 \includegraphics[width=6.0 cm, angle=270]{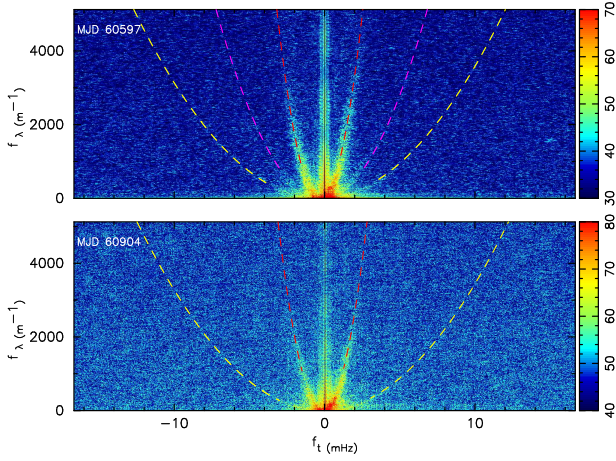}
    \caption{The secondary spectra of PSR~B1257+12 from observations on MJDs~60597 and 60904. The dashed lines indicate the best-fit results of the arc curvature for the inner (red), middle (pink), and outer arcs (yellow).}
\label{fig:two_ss}
\end{figure}

In the secondary spectra, throughout these observations of PSR~B1257+12, a maximum of three arcs were detected on MJDs~59809, 59839, and 60597: the inner, middle, and outer arcs. As mentioned in Section~\ref{sec:Obs}, we performed a tracking observation with a maximum duration of 5 hours on MJD~60904 using FAST, during which only the inner and outer arcs were detected. As shown in Figure~\ref{fig:two_ss}, three arcs are present on MJD~60597, while only two arcs are detected on MJD~60904. The inner arc is the strongest, followed by the outer arc, with the middle arc being the weakest. As shown in Figure~\ref{fig:DS_four} in Section~\ref{sec:DS_ACF}, periodic fringe patterns are observed in the dynamic spectra at MJDs 60683 and 60797. We inspected the corresponding secondary spectra but found no clear arclets or isolated patch-like structures. This may indicate that the sub-images responsible for producing these features are too weak to be detected. Following \cite{2021NatAs...5..788Y}, since the peaks of the arcs are offset from the coordinate origin due to the phase gradient, especially for the inner arc, we describe the arcs using,
\begin{equation}
f_\lambda=\eta f^2_t+\gamma f_t
\label{equ:arc}
\end{equation}
where $\eta$ is the arc curvature, and $\gamma$ is the gradient parameter. 

\begin{figure}[!h]
\center
 \includegraphics[width=6.0 cm, angle=270]{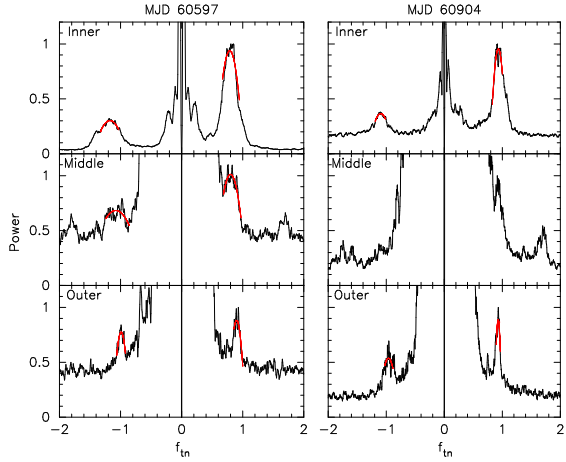}
    \caption{The Doppler profiles of the three arcs observed on MJDs~60597 and 60904. From the top to bottom panels, the red lines represent the parabolic fits to the arc peaks corresponding to the inner, middle, and outer arcs.}
\label{fig:two_DP}
\end{figure}

To estimate $\eta$ and $\gamma$, we followed the procedure described in \cite{2020ApJ...904..104R}. First, we transformed the secondary spectra for the inner, middle, and outer arcs using arc curvatures $\eta_0$ of 600, 100, and 30~m$^{-1}$~mHz$^{-2}$, respectively, so that the scintillation arcs appear as vertical features in the normalized secondary spectra. Second, we averaged the power along $f_\lambda$ to obtain the power distribution as a function of the normalized Doppler frequency, $f_{\rm tn}$. For the secondary spectra shown in Figure~\ref{fig:two_ss}, the corresponding power distributions, referred to as the Doppler profiles, are presented in Figure~\ref{fig:two_DP}. In the left panels, the Doppler profiles on MJD~60597 for the inner, middle, and outer arcs all show double peaks, indicating that three distinct arcs are detected on this epoch, whereas only two pronounced arcs are detected on MJD~60904.
Third, as shown in Figure~\ref{fig:two_DP}, we fitted the peak regions of the Doppler profiles with a parabolic function and obtained $\beta_{N}$ and $\beta_{P}$ for the negative- and positive-$f_{\rm tn}$ sides, respectively. For the inner and outer arcs, the measured values and 1$\sigma$ uncertainties of $\beta_{N}$ and $\beta_{P}$ are listed in columns 5 and 6 of Tables~\ref{tab:inner} and \ref{tab:outer}, respectively. For the middle arcs, they are given in columns 4 and 5 of Table~\ref{tab:middle}. Fourth, following the steps described in \cite{yky+26}: (i) separately computing $\eta_0/\beta_N^2$ and $\eta_0/\beta_P^2$ for the left and right sides of each arc, (ii) constructing the corresponding left and right branches of the arcs, and (iii) fitting the arcs with Equation~\ref{equ:arc} to obtain $\eta$ and $\gamma$, we got the final measurements of $\eta$ and $\gamma$ for the inner, middle, and outer arcs. For the inner and outer arcs, the resulting values and their 1$\sigma$ uncertainties are listed in columns 7 and 8 of Tables~\ref{tab:inner} and \ref{tab:outer}, respectively. For the middle arcs, they are listed in columns 6 and 7 of Table~\ref{tab:middle}. In total, we obtained 31 detections for the inner arc, 3 detections for the middle arc, and 15 detections for the outer arc.

From Figures~\ref{fig:two_ss} and \ref{fig:two_DP}, for the inner arc, we note that the arc power on the positive Doppler side is brighter than that on the negative Doppler side. Following the discussion and simulations presented in \citet{2006ApJ...637..346C}, refractive effects can produce two distinct observational signatures in the secondary spectrum. First, a refractive phase gradient can shift the scattered image, leading to an offset of the arc apex, consistent with the shifted arc structures observed in our measurements. Second, refractive focusing and defocusing, together with asymmetric scattering brightness distributions, can enhance the scattered power on one side of the Doppler spectrum, thereby producing asymmetric arc brightness.

\begin{table*}
\caption{Value and uncertainty of $\xi$, $\beta_{N}$, $\beta_{P}$, $\eta$, $\gamma$, $\theta_g$ and the change rate of the DM for the inner arc.}
\centering
\setlength\tabcolsep{2.0pt}
\footnotesize
\begin{tabular}{cccccccccc}
\hline
  UTC& MJD & Time & $\xi$& $\beta_{N}$& $\beta_{P}$& $\eta_{\rm inner}$ & $\gamma_{\rm inner}$ & $\theta_g$&$\Delta DM$/AU~(10$^{-5}$)\\
  \hline
       &  &  min & &  &  & m$^{-1}$~mHz$^{-2}$ &  m$^{-1}$~mHz$^{-1}$ &$\mu$as & pc~cm$^{-3}$~AU$^{-1}$\\
  \hline
2022/08/18& 59809& 180&$-$2.22$\pm$0.04& $-$1.020$\pm$0.002& 0.992$\pm$0.002& 592.9$\pm$1.8& 27.8$\pm$0.3   & 12.0$\pm$0.5 & $-$1.10$\pm$0.05\\
2022/09/06& 59828&  30 & & $-$1.007$\pm$0.004& 0.889$\pm$0.003& 668.9$\pm$3.8 & 132.1$\pm$1.5 & 54.5$\pm$2.7 & $-$4.79$\pm$0.25\\
2022/09/08& 59830&  30& & $-$1.080$\pm$0.006& 0.833$\pm$0.003& 660.8$\pm$5.1 & 271.4$\pm$3.3 & 111.7$\pm$5.7& $-$10.19$\pm$0.52\\
2022/09/10& 59832&  30& & $-$0.997$\pm$0.003& 0.865$\pm$0.005& 693.9$\pm$4.6 & 153.3$\pm$1.8 & 62.9$\pm$3.2 & $-$5.74$\pm$0.29\\
2022/09/12& 59834&  30& & $-$1.015$\pm$0.005& 0.856$\pm$0.005& 687.9$\pm$5.8 & 182.8$\pm$2.1 & 74.8$\pm$3.9 & $-$6.83$\pm$0.35\\
2022/09/14& 59836&  30& & $-$0.765$\pm$0.015& 0.991$\pm$0.009& 784.2$\pm$16.2 & $-$294.7$\pm$3.6&$-$120.4$\pm$6.3& 11.0$\pm$0.6\\
2022/09/16& 59838&  30& & $-$0.840$\pm$0.007& 0.853$\pm$0.008& 837.4$\pm$10.5 & $-$18.3$\pm$0.2&$-$7.5$\pm$0.4 & 0.68$\pm$0.04\\
2022/09/17& 59839& 180&$-$1.97$\pm$0.04&$-$1.041$\pm$0.004& 0.865$\pm$0.006& 663.3$\pm$5.4 & 194.9$\pm$2.4 & 79.4$\pm$4.2 & $-$7.24$\pm$0.38\\
2022/09/18& 59840&  30& & $-$0.937$\pm$0.005& 0.941$\pm$0.004& 680.5$\pm$4.6 & $-$4.5$\pm$0.1&$-$1.8$\pm$0.1 & 0.17$\pm$0.01\\
2022/09/20& 59842&  30& & $-$0.957$\pm$0.004& 0.888$\pm$0.002& 705.5$\pm$3.6 & 81.6$\pm$0.9  & 33.2$\pm$1.8 & $-$3.03$\pm$0.16\\
2022/09/22& 59844&  30& & $-$0.940$\pm$0.007& 0.912$\pm$0.003& 699.8$\pm$5.8 & 32.9$\pm$0.3  & 13.4$\pm$0.7 & $-$1.22$\pm$0.06\\
2022/09/24& 59846&  30& & $-$0.971$\pm$0.004& 0.847$\pm$0.006& 727.8$\pm$6.0 & 151.0$\pm$1.8 & 61.4$\pm$3.3 & $-$5.60$\pm$0.30\\
2022/09/26& 59848&  30& & $-$0.983$\pm$0.012& 0.872$\pm$0.004& 698.7$\pm$9.1 & 129.9$\pm$1.5 & 52.8$\pm$2.8 & $-$4.81$\pm$0.26\\
2022/09/29& 59851&  30& & $-$0.942$\pm$0.003& 0.898$\pm$0.003& 725.0$\pm$3.5 & 77.8$\pm$0.9  & 31.6$\pm$1.7 & $-$2.89$\pm$0.16\\
2022/10/01& 59853&  30& & $-$0.979$\pm$0.013& 0.838$\pm$0.003& 729.9$\pm$11.0 & 173.3$\pm$2.1& 70.6$\pm$3.8 & $-$6.43$\pm$0.34\\
2022/10/03& 59855&  30& & $-$1.082$\pm$0.004& 0.835$\pm$0.007& 658.0$\pm$6.1 & 270.3$\pm$3.3 & 110.2$\pm$5.9 & $-$10.05$\pm$0.54\\
2022/10/05& 59857&  30& & $-$0.999$\pm$0.004& 0.825$\pm$0.003& 724.5$\pm$4.4 & 210.5$\pm$2.4 & 86.0$\pm$4.6 & $-$7.84$\pm$0.42\\
2022/10/11& 59863& 180&$-$2.06$\pm$0.04&$-$1.117$\pm$0.001& 0.836$\pm$0.003& 635.7$\pm$2.7& 295.3$\pm$3.9  & 121.6$\pm$6.5 & $-$11.09$\pm$0.59\\
2022/11/01& 59884&  60& & $-$1.182$\pm$0.008& 0.828$\pm$0.003& 601.7$\pm$5.8& 351.3$\pm$4.8  &  151.9$\pm$7.2 & $-$13.86$\pm$0.66\\
2022/12/31& 59944&  60& & $-$1.570$\pm$0.004& 0.943$\pm$0.003& 387.8$\pm$2.4& 392.5$\pm$6.6  & 216.2$\pm$5.4 & $-$19.72$\pm$0.49\\
2024/10/14& 60597& 210& $-$2.51$\pm$0.03&$-$1.172$\pm$0.002& 0.788$\pm$0.006& 634.2$\pm$5.0 & 399.9$\pm$5.4 & 165.6$\pm$8.7 & $-$15.10$\pm$0.79\\
2024/11/23& 60637& 180 &$-$2.21$\pm$0.04& $-$1.136$\pm$0.004& 1.018$\pm$0.003& 518.0$\pm$2.6 & 102.4$\pm$1.2 & 48.1$\pm$1.8 & $-$4.38$\pm$0.17\\
2025/01/08& 60683& 210&$-$2.09$\pm$0.04&$-$1.226$\pm$0.003& 1.329$\pm$0.007& 367.9$\pm$2.3 & $-$63.5$\pm$0.6&36.2$\pm$0.6 & $-$3.30$\pm$0.06\\
2025/02/27& 60733& 210& $-$2.33$\pm$0.03& $-$1.781$\pm$0.004& 1.294$\pm$0.002& 256.6$\pm$1.0 & 206.8$\pm$2.7  &135.5$\pm$1.8 & $-$12.36$\pm$0.17\\
2025/03/28& 60762& 120& $-$1.97$\pm$0.04& $-$1.972$\pm$0.009& 1.229$\pm$0.003& 238.6$\pm$1.9 & 288.2$\pm$4.5  &192.5$\pm$3.1 & $-$17.56$\pm$0.28\\
2025/04/23& 60788& 120& $-$2.00$\pm$0.05&$-$1.550$\pm$0.003& 1.451$\pm$0.002& 266.6$\pm$0.7 & 44.3$\pm$0.6   &28.9$\pm$0.4 & $-$2.63$\pm$0.04\\
2025/05/02& 60797& 120&$-$2.30$\pm$0.05& $-$1.367$\pm$0.002& 1.562$\pm$0.002& 280.4$\pm$0.6 & $-$91.5$\pm$0.9&$-$58.6$\pm$0.6& $-$5.34$\pm$0.06\\
2025/05/14& 60809& 120&$-$1.87$\pm$0.05&$-$1.500$\pm$0.002& 1.336$\pm$0.004& 298.9$\pm$1.0 & 82.1$\pm$0.9   & 51.1$\pm$0.6 & $-$4.66$\pm$0.05\\
2025/07/17& 60873& 180&$-$2.37$\pm$0.04&$-$1.228$\pm$0.001& 0.994$\pm$0.006& 488.6$\pm$3.0 & 190.7$\pm$2.1  & 93.2$\pm$2.6 & $-$8.50$\pm$0.24\\
2025/07/22& 60878& 120&$-$2.29$\pm$0.05&$-$1.206$\pm$0.002& 0.974$\pm$0.002& 507.7$\pm$1.7 & 196.4$\pm$2.1  &93.9$\pm$2.9 & $-$8.57$\pm$0.26\\
2025/08/17& 60904& 288& $-$1.91$\pm$0.03&$-$1.095$\pm$0.002& 0.932$\pm$0.003& 585.9$\pm$2.4 &159.7$\pm$1.8   & 69.3$\pm$3.0 & $-$6.32$\pm$0.27\\
\hline
   \end{tabular}
   \label{tab:inner}
\end{table*}

\begin{table*}
\caption{Value and uncertainty of $\beta_{N}$, $\beta_{P}$, $\eta$, $\gamma$, $\theta_g$ and the change rate of the DM for the middle arc.}
\centering
\setlength\tabcolsep{2.0pt}
\footnotesize
\begin{tabular}{ccccccccc}
\hline
  UTC& MJD & Time & $\beta_{N}$& $\beta_{P}$& $\eta_{\rm middle}$ & $\gamma_{\rm middle}$ & $\theta_g$&$\Delta DM$/AU~(10$^{-5}$)\\
  \hline
       & & min  &  &  & m$^{-1}$~mHz$^{-2}$ &  m$^{-1}$~mHz$^{-1}$ &$\mu$as & pc~cm$^{-3}$~AU$^{-1}$\\
  \hline
2022/08/18& 59809& 180& $-$1.004$\pm$0.005& 0.955$\pm$0.002& 104.26$\pm$0.59& 20.98$\pm$0.24& 14.5$\pm$0.2& $-$1.32$\pm$0.02\\
2022/09/17& 59839& 180& $-$1.025$\pm$0.004& 0.933$\pm$0.002& 104.45$\pm$0.51& 39.43$\pm$0.45& 27.0$\pm$0.3& $-$2.46$\pm$0.03\\
2024/10/14& 60597& 210& $-$1.078$\pm$0.011& 0.801$\pm$0.004& 104.03$\pm$0.41& 40.13$\pm$0.45& 27.6$\pm$0.3& $-$2.52$\pm$0.03\\
  \hline
   \end{tabular}
   \label{tab:middle}
\end{table*}

\begin{table*}
\caption{Value and uncertainty of $\xi$, $\beta_{N}$, $\beta_{P}$, $\eta$, $\gamma$, $\theta_g$ and the change rate of the DM for the outer arc.}
\centering
\setlength\tabcolsep{2.0pt}
\footnotesize
\begin{tabular}{cccccccccc}
\hline
  UTC & MJD & Time & $\xi$& $\beta_{N}$& $\beta_{P}$& $\eta_{\rm outer}$ & $\gamma_{\rm outer}$ & $\theta_g$&$\Delta DM$/AU~(10$^{-5}$)\\
  \hline
       & & min & &  &  & m$^{-1}$~mHz$^{-2}$ &  m$^{-1}$~mHz$^{-1}$ &$\mu$as & pc~cm$^{-3}$~AU$^{-1}$\\
  \hline
2022/08/18& 59809& 180&$-1.81\pm0.08$&$-$0.927$\pm$0.001& 1.026$\pm$0.002& 31.5$\pm$0.3& $-$23.62$\pm$0.27& $-$16.9$\pm$0.2& 1.54$\pm$0.02\\
2022/09/06& 59828&  30& & $-$0.967$\pm$0.006& 0.972$\pm$0.005& 31.9$\pm$0.3& $-$1.20$\pm$0.03& $-$0.86$\pm$0.02& 0.078$\pm$0.002\\
2022/09/17& 59839& 180& $-1.19\pm0.07$&$-$0.945$\pm$0.002& 1.002$\pm$0.003& 31.7$\pm$0.1&$-$13.54$\pm$0.15& $-$9.68$\pm$0.11& 0.88$\pm$0.01\\
2022/09/22& 59844&  30& & $-$0.818$\pm$0.007& 1.029$\pm$0.007& 35.4$\pm$0.4& $-$55.63$\pm$0.66&$-$39.65$\pm$0.48& 3.62$\pm$0.04\\
2022/10/05& 59857&  30& & $-$0.959$\pm$0.005& 0.951$\pm$0.017& 32.9$\pm$0.6&1.97$\pm$0.03&1.41$\pm$0.02& $-$0.128$\pm$0.002\\
2022/10/11& 59863& 180&$-1.70\pm0.08$&$-$1.013$\pm$0.002& 0.935$\pm$0.002& 31.6$\pm$0.1& 18.50$\pm$0.21& 10.38$\pm$0.06& $-$0.95$\pm$0.01\\
2024/10/14& 60597& 210&$-2.25\pm0.07$&$-$0.990$\pm$0.003& 0.901$\pm$0.006&33.6$\pm$0.2 & 22.4$\pm$0.24& 16.03$\pm$0.17& $-$1.46$\pm$0.02\\
2024/11/23& 60637& 180&$-1.16\pm0.08$&$-$0.990$\pm$0.007& 0.982$\pm$0.008& 30.9$\pm$0.3 & 1.85$\pm$0.02& 1.34$\pm$0.01 & $-$0.122$\pm$0.001\\
2025/01/08& 60683& 210&$-1.98\pm0.07$&$-$0.988$\pm$0.004& 1.003$\pm$0.005& 30.3$\pm$0.2 & $-$3.41$\pm$0.03& $-$2.51$\pm$0.02& 0.229$\pm$0.002\\
2025/02/27& 60733& 210& $-1.74\pm0.07$&$-$1.104$\pm$0.005& 0.896$\pm$0.017& 30.2$\pm$0.5 & 46.61$\pm$0.54& 34.79$\pm$0.41& $-$3.17$\pm$0.04\\
2025/05/02& 60797& 120& $-2.10\pm0.08$&$-$0.957$\pm$0.001& 1.016$\pm$0.002& 30.8$\pm$0.1 & $-$13.64$\pm$0.15 &$-$10.12$\pm$0.11&0.92$\pm$0.01\\
2025/05/14& 60809& 120&$-2.22\pm0.08$ & $-$1.029$\pm$0.007& 0.953$\pm$0.001& 30.6$\pm$0.1 & 17.42$\pm$0.18& 12.88$\pm$0.13& $-$1.17$\pm$0.01\\
2025/07/17& 60873& 180& $-2.38\pm0.08$&$-$1.050$\pm$0.003& 0.869$\pm$0.003& 32.7$\pm$0.1 & 44.23$\pm$0.51 & 31.97$\pm$0.37& $-$2.91$\pm$0.03\\
2025/07/22& 60878& 120& $-1.82\pm0.10$&$-$1.035$\pm$0.001& 0.929$\pm$0.002& 31.2$\pm$0.1 & 24.74$\pm$0.27 &17.86$\pm$0.19& $-$1.63$\pm$0.02\\
2025/08/17& 60904& 288&$-1.67\pm0.04$ &$-$0.962$\pm$0.009& 0.929$\pm$0.002& 33.6$\pm$0.3 &8.31$\pm$0.09 & 5.95$\pm$0.07& $-$0.54$\pm$0.01\\ 
  \hline
   \end{tabular}
   \label{tab:outer}
\end{table*}

\section{The scattering screen and DM variations}\label{sec:SL_DM}
In this section, we present the locations and turbulence spectral indices of the scattering screens, as well as the DM variation rates inferred from the three detected arcs. Because the curvature of the inner arc exhibits a clear modulation caused by Earth’s annual motion, we can accurately determine the location of the scintillation screen from the annual variation of the arc curvature. Because the inner arc corresponds to a scattering screen located closer to the Earth, we define its location using its distance from the Earth. For the middle and outer arcs, we define their locations using their distances from the pulsar.

\subsection{The locations of these three scattering screens}\label{sec:location}
\begin{table}
\caption{The fitted values and uncertainties of $s_{\rm inner}$, $\psi$, $V_{\rm scr,\delta}$, and $V_{\rm scr,\delta}$ under both the isotropic and the anisotropic cases, as well as the corresponding reduced chi-square values.}
\centering
\setlength\tabcolsep{1.5pt}
\footnotesize
\begin{tabular}{cccccc}
\hline
Case& $s_{\rm inner}$ & $\psi$ & $V_{\rm scr, \alpha}$ & $V_{\rm scr,\delta}$ & $\chi^2_\nu$ \\
    &   & deg & km~s$^{-1}$ & km~s$^{-1}$ &  \\
  \hline
I&0.671$\pm$0.011& &$-$7.16$\pm$2.16& $-$41.07$\pm$5.69&40.1\\
II&0.677$\pm$0.013&$-$45.75$\pm$3.78& $-$7.80& $-$39.02&40.5\\
\hline
   \end{tabular}
   \label{tab:fit}
\end{table}

\begin{figure}[!h]
\center
 \includegraphics[width=7.0 cm, angle=270]{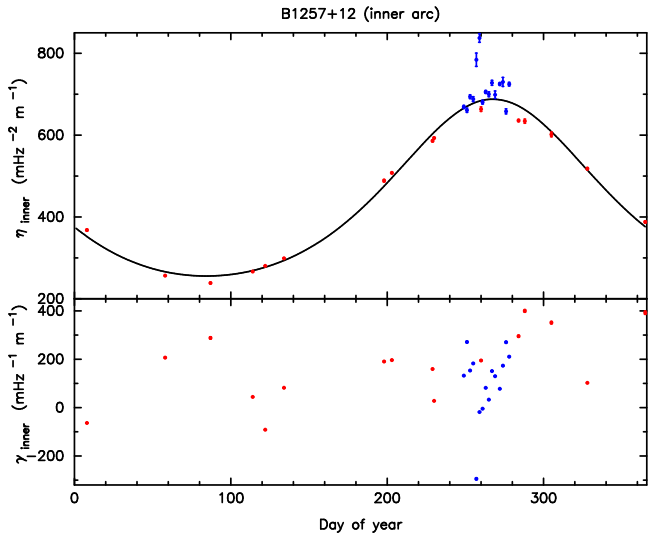}
    \caption{The measured $\eta$ and $\gamma$ for the inner arc as a function of day of year. The blue points represent observations with a duration of 30~min, while the red points correspond to all observations longer than 30~min. In the upper panel, the solid line show the best-fit results for Case I listed in Table~\ref{tab:fit}.}
\label{fig:eta_inner}
\end{figure}

From \cite{2021NatAs...5..788Y} and \cite{2020ApJ...904..104R}, in Equation~\ref{equ:arc}, 
\begin{equation}
\eta=3.08567758\times10^7\frac{D_{\rm kpc}s(1-s)}{2V^2_{\rm eff, \parallel}}, 
\label{eq:eta_D}
\end{equation}
\begin{equation}
\gamma=3.08567758\times10^{13}\frac{D_{\rm kpc}(1-s)\theta_g}{\lambda_cV_{\rm eff, \parallel}},
\label{eq:gamma_D}
\end{equation}
here, $\eta$ is in units of m$^{-1}$ mHz$^{-2}$, $\gamma$ is in m$^{-1}$ mHz$^{-1}$, and $\theta_g$ is the refractive angle in radians. The $V_{\rm eff,\parallel}$ (in km s$^{-1}$) is the effective velocity component parallel to the anisotropy axis and under isotropic scattering is given by 
\begin{equation}
V_{\rm eff,\parallel} = \left|(1-s) \vec{V}_{\rm pulsar,\perp}  + s \vec{V}_{{\rm Earth, \perp}} - \vec{V}_{\rm scr, \perp}\right|.
\label{eq:V_I}  
\end{equation}
where $\vec{V}_{\rm pulsar,\perp}$ is the pulsar transverse velocity obtained from the measured proper motion, $\vec{V}_{{\rm Earth}}$ is the component of the Earth’s velocity perpendicular to the line of sight to the pulsar and $\vec{V}_{\rm scr, \perp}$ is the component of the velocity of the scattering screen perpendicular to the line of sight. And under anisotropic scattering, following \cite{mnc+23}, $V_{\rm eff,\parallel}$ is given by,  
\begin{multline}
V_{\rm eff, \parallel} = \Big| 
  \big[(1-s) V_{\rm pulsar,\alpha} + sV_{\rm Earth,\alpha}-V_{\rm scr,\alpha}\big] \sin\psi \\
  + \big[(1-s) V_{\rm pulsar,\delta} + sV_{\rm Earth,\delta}-V_{\rm scr,\delta}\big] \cos\psi
\Big|
\label{eq:V_II}
\end{multline}
where the anisotropy axis is oriented at an angle $\psi$ measured anti-clockwise from the positive declination direction. For PSR~B1257+12, we adopt a pulsar distance of $D = 710 \pm 40$ pc in the following calculations. We use the proper motions measured from timing, $\mu_{\alpha} = 45.50 \pm 0.04$ mas yr$^{-1}$ and $\mu_{\delta} = -84.70 \pm 0.07$ mas yr$^{-1}$ \citep{2003ApJ...591L.147K}. These correspond to transverse velocities of $V_{\rm pulsar,\alpha} = 153 \pm 9$ km s$^{-1}$ and $V_{\rm pulsar,\delta} = -285 \pm 16$ km s$^{-1}$ in the right ascension and declination directions, respectively.

In Figure~\ref{fig:eta_inner}, we show the measured values of $\eta_{\rm inner}$ and $\gamma_{\rm inner}$ for the inner arc as a function of day of year. Compared with $\gamma_{\rm inner}$, the distribution of $\eta_{\rm inner}$ exhibits a clear modulation caused by Earth’s annual motion. From Equations~\ref{eq:eta_D} and \ref{eq:gamma_D}, we note that, in contrast to the variation in $\eta_{\rm inner}$, the variation in $\gamma_{\rm inner}$ is affected not only by factors associated with Earth’s annual motion, but also by changes in $\theta_g$ over the observation epochs. As a result, its annual modulation is less pronounced. Therefore, in the following analysis, we use the annual modulation of $\eta_{\rm inner}$ to determine the location of the corresponding scattering screen. 

As shown in the top panel of Figure~\ref{fig:eta_inner} and Table~\ref{tab:fit}, by excluding the two data sets with $\eta_{\rm inner} > 700$~mHz$^{-2}$~m$^{-1}$, which clearly deviate from the annual modulation trend, we fitted the measured $\eta_{\rm inner}$ under two cases: Cases I and II. In Case I, we assume isotropic scattering and fit for $s_{\rm inner}$, $V_{\rm scr,\alpha}$ and $V_{\rm scr,\delta}$, where the scattering screen is located at a distance $s_{\rm inner}D_{\rm kpc}$ from the pulsar. In Case II, we assume anisotropic scattering and fit for $s_{\rm inner}$, $V_{\rm scr,\alpha}$, $V_{\rm scr,\delta}$ and $\psi$, where $\psi$ is defined under Equation~\ref{eq:V_II}. Under the assumption of isotropic scattering (Case I in Table~\ref{tab:fit}), we obtained the transverse velocity of the scattering screen and found its distance from the Earth to be $D_{\rm se, inner}=233\pm28$~pc. In Figure~\ref{fig:eta_inner}, the Case I fitting result is shown as a solid line, which is in good agreement with the data points. Compared with Case I, in Case II, all fitted parameters except $\psi$ are consistent with the Case I values within their 1-$\sigma$ uncertainties. However, due to the strong covariance between $\psi$ and $V_{\rm scr,\alpha}$ and $V_{\rm scr,\delta}$, the fitted uncertainties of $V_{\rm scr,\alpha}$ and $V_{\rm scr,\delta}$ are very large. Therefore, their corresponding uncertainties are not listed in the table. Because the Earth’s velocity is small compared to that of the pulsar, from Table~\ref{tab:fit}, we find that in Case II the anisotropy axis of the pulsar’s scattered image is essentially aligned with $V_{\rm eff,\parallel}$, showing no significant difference from the isotropic case. 

\begin{figure}[!h]
\center
 \includegraphics[width=6.5 cm, angle=270]{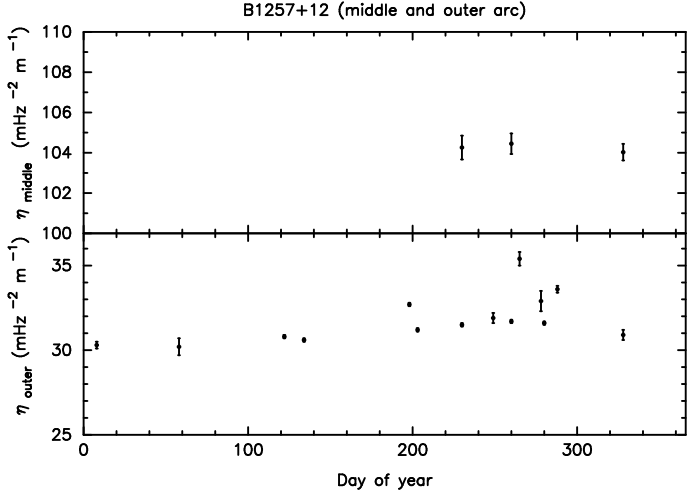}
    \caption{The measured arc curvature as a function of day of year for the middle and outer arcs.}
\label{fig:eta_middle_outer}
\end{figure}

From Figure~\ref{fig:eta_middle_outer}, the curvatures of both the middle and outer arcs exhibit virtually no modulation due to Earth’s motion. In the following calculations, we therefore neglect both the Earth’s velocity and the screen velocity, and use the measured mean values of $\eta_{\rm middle}$ and $\eta_{\rm outer}$ to derive the screen–to–pulsar distances for the middle and outer arcs. For the middle arc, the three available measurements of $\eta_{\rm middle}$ are consistent within their 1-$\sigma$ uncertainties, yielding an average value of 104.25$\pm$0.52~m$^{-1}$~mHz$^{-2}$. In contrast to the middle arc, the 15 available measurements of $\eta_{\rm outer}$ exhibit substantially larger variations. This could be attributed to variations in the screen velocity, or it may suggest that the measurement uncertainties of $\eta_{\rm outer}$ were underestimated. Consequently, we adopt the mean value of $\eta_{\rm outer}$, 31.9$\pm$0.2~m$^{-1}$~mHz$^{-2}$, to calculate the corresponding screen–to–pulsar distance. Compared to the inner arc, the middle and outer arcs exhibit weaker scintillation. Based on the simulations in \cite{2020ApJ...904..104R} (see their Figure 6), in the weak scintillation regime, the arc curvature is largely insensitive to variations in the anisotropic scattering angle. Therefore, under the assumption of isotropic scattering, reliable estimates of the scattering screen locations can still be obtained for both the middle and outer arcs. Following \cite{2021NatAs...5..788Y}, and assuming Gaussian distributions for $D$, $\mu_{\alpha}$, $\mu_{\delta}$, and $\eta$, we use a Monte Carlo analysis to obtain the distributions of $D_{\rm sp}=sD$ for both the middle and outer arcs. Assuming isotropic scattering, this gives $s_{\rm middle} = 0.499 \pm 0.014$ and $D_{\rm sp,middle} = 354 \pm 22$~pc for the middle arc, and $s_{\rm outer} = 0.234 \pm 0.010$ and $D_{\rm sp,outer} = 166 \pm 12$~pc for the outer arc, respectively. 

\subsection{The turbulence of the scattering screen}
\begin{figure*}[!h]
\center
 \includegraphics[width=10.0 cm, angle=270]{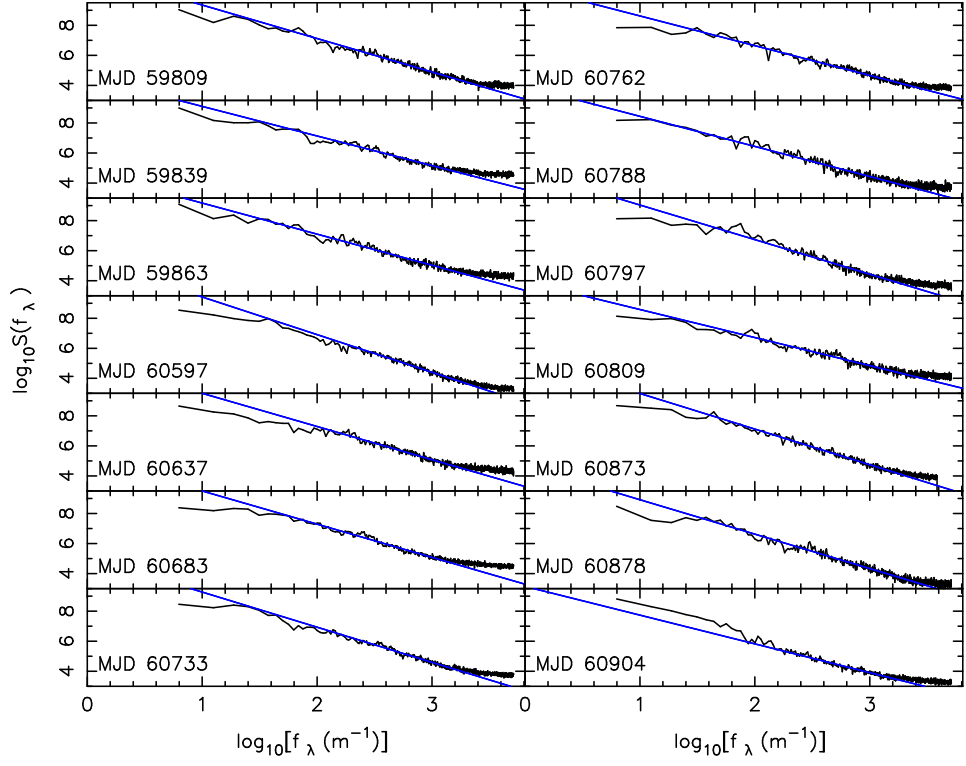}
    \caption{The delay profiles for inner arcs at 1250~MHz for all observations with durations of at least 120~min. The black lines represent the observed delay profiles, while the blue lines show the best-fit slopes to the linear portions, corresponding to the $f_\lambda$ range between 200 and 1800~m$^{-1}$, i.e., an X-axis range from 2.3 to 3.2.}
\label{fig:sf_14}
\end{figure*}

\begin{figure*}[!h]
\center
 \includegraphics[width=10.0 cm, angle=270]{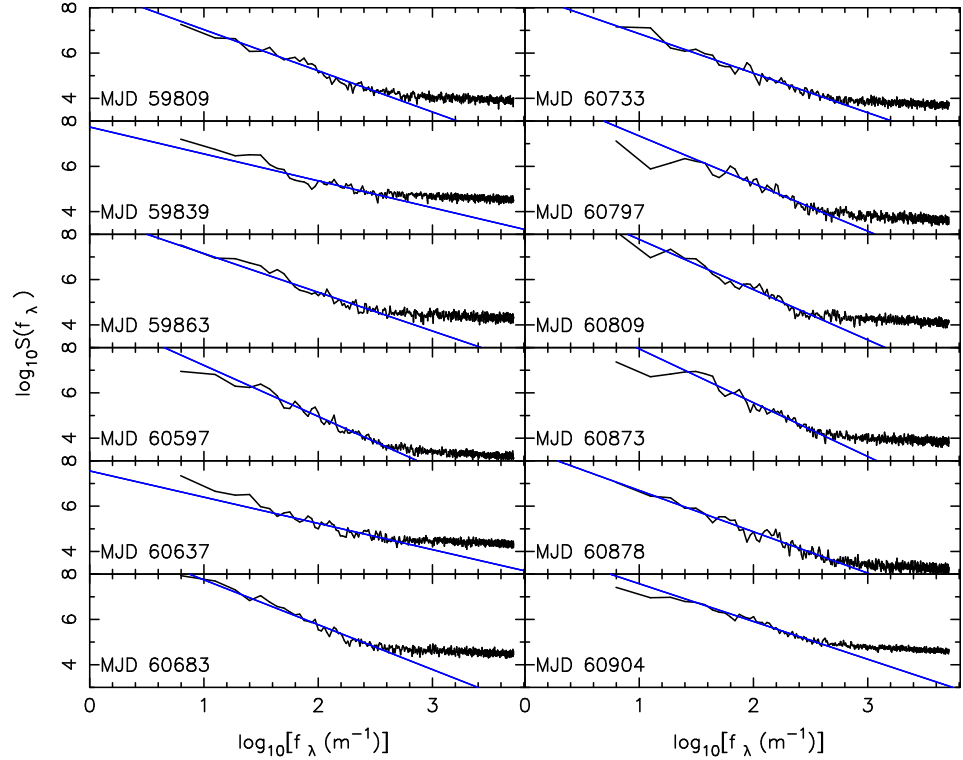}
    \caption{The delay profiles for outer arcs at 1250~MHz for all observations with durations of at least 120~min. The black lines represent the observed delay profiles, while the blue lines show the best-fit slopes to the linear portions, corresponding to the $f_\lambda$ range between 32 and 398~m$^{-1}$, i.e., an X-axis range from 1.5 to 2.6.}
\label{fig:sf_12}
\end{figure*}

Because the inner and outer arcs are stronger and more frequently detected, we used the normalized secondary spectra from 14 observations (inner arc) and 12 observations (outer arc), each with durations of at least 120 min, to estimate the spectral exponent of the associated scattering turbulence. Following \cite{2022ApJ...939...75Y}, we obtained the delay profiles by summing over the inner arc regions in the normalized secondary spectra, as shown in the upper and lower panels of Figure~\ref{fig:two_DP}. The resulting delay profiles for the inner and outer arcs are presented in Figures~\ref{fig:sf_14} and \ref{fig:sf_12}, respectively. Following Appendix B of \cite{2020ApJ...904..104R}, in the weak scintillation regime with isotropic scattering, the arc power scales as
\begin{equation}
S(f_\lambda, f_t)\propto f_\lambda^{-(\alpha+1)/2}\frac{1}{\sqrt{1-f_{\rm tn}^2}}.
\end{equation}
For Kolmogorov turbulence ($\alpha=11/3$), this yields
\begin{equation}
\xi = \frac{-11/3 - 1}{2} = -\frac{7}{3}.
\end{equation}
Although this relation was originally derived for the idealized case of weak and isotropic scintillation, previous studies have shown that observed arc power distributions can remain broadly consistent with similar power-law behavior even in moderate scattering regimes \citep{2022ApJ...939...75Y, 2024MNRAS.527.7568O}. Therefore, here we use this relation primarily as an approximate phenomenological description of the observed delay-profile behavior. For these delay profiles, we estimated the delay-profile exponent, $\xi$, by fitting the slopes of the linear portions of the curves. For the inner arc, the fit was performed over the $f_\lambda$ range of 200--1800~m$^{-1}$ (corresponding to an x-axis range of 2.3--3.2), while for the outer arc, the range is 32--398~m$^{-1}$ (i.e., an x-axis range of 1.5--2.6). The best-fit slopes, i.e. the exponents, are listed in column 4 of Tables~\ref{tab:inner} and \ref{tab:outer} and are shown as blue lines in Figures~\ref{fig:sf_14} and \ref{fig:sf_12}. For the inner arc, we found that, except for the observation on MJD~60597, the estimated spectral exponents are either consistent with, or smaller than, the Kolmogorov value. Similarly, for the outer arc, the estimated spectral exponents are also consistent with or below the Kolmogorov expectation. Using Equation~2 of \cite{2020ApJ...904..104R} and the value $s_{\rm inner}=0.671\pm0.011$ (Section~\ref{sec:location}), the receive angles $\theta$ corresponding to $f_\lambda=200$ and $1800~\mathrm{m^{-1}}$ are $4.4\times10^{-4}$~$\mu$as and $4.0\times10^{-3}$~$\mu$as, respectively. Applying Equation (2.4) of \cite{1990ARA&A..28..561R}, the corresponding spatial scales in the scattering plasma range from $2.0\times10^9$ to $1.8\times10^{10}$~km. For the outer arc, using $s_{\rm outer}=0.234\pm0.010$ (Section~\ref{sec:location}), the receive angles corresponding to $f_\lambda=32$ and $398~\mathrm{m^{-1}}$ are $1.8\times10^{-7}$~$\mu$as and $2.3\times10^{-6}$~$\mu$as. These correspond to spatial scales ranging from $5.9\times10^{10}$ to $7.4\times10^{11}$~km.

\subsection{The DM variations of PSR~B1257+12}
Following the same procedure used to estimate $D_{\rm sp}$, we obtained $\theta_g$ for the inner, middle and outer arcs using Equation~\ref{eq:gamma_D} together with the measured values of $\gamma$. For the inner and outer arcs, the resulting $\theta_g$ values are presented in column 9 of Tables~\ref{tab:inner} and \ref{tab:outer}, while for the middle arc, the resulting $\theta_g$ values are presented in column 8 of Table~\ref{tab:middle}. According to \cite{2023MNRAS.521.6392R}, for each epoch, the phase gradient across the scattering disk is given by
\begin{equation}
\nabla\phi_\parallel (t_i)=\theta_g*k=\theta_g*\frac{2\pi}{\lambda},
\end{equation}
and the corresponding change rate of DM is
\begin{equation}
\rm Rate(DM)=-3.84*10^{-8}\nu\nabla\phi_\parallel(t_i)*AU_{\rm km}
\label{eq:dm}
\end{equation}
where Rate(DM) is in pc~cm$^{-3}$~AU$^{-1}$ and $\nu$ is in GHz. Based on Equation~\ref{eq:dm}, we calculated the DM change rates for the inner, middle, and outer arcs, which are listed in column 10 of Tables~\ref{tab:inner} and \ref{tab:outer}, and in column 9 of Table~\ref{tab:middle}. 

Based on timing analyses, Lei et al. (in prep.) found that the DM of PSR~B1257+12 is gradually decreasing. For the inner arc, approximately 90\% of the derived DM change rates, Rate(DM), are negative, consistent with the DM trend inferred from timing. For the middle arc, all measured Rate(DM) values are negative, which is also in agreement with the DM decrease seen in the timing analysis. However, for the outer arc, nearly half of the measured DM change rates are positive, indicating that the overall DM variation of PSR~B1257+12 is not dominated by the scattering screen associated with the outer arc, which is located closer to the pulsar. In addition, the average DM change rate inferred for the outer arc of 1.3$\pm$1.1~(10$^{-5}$)~pc~cm$^{-3}$~AU$^{-1}$ is only about 18\% of that derived for the inner arc of 7.2$\pm$4.9~(10$^{-5}$)~pc~cm$^{-3}$~AU$^{-1}$. These results indicate that the region surrounding PSR~B1257+12 is relatively free of dense ionized material.

\section{Summary and conclusions}\label{sec:SD}
In this paper, we investigate the scintillation properties of the first known pulsar planetary system using high-sensitivity FAST observations. A total of 31 observations with durations of at least 30~min were analysed. For the 14 observations lasting longer than 120~min, the 1D ACF analysis was applied, yielding measurements of the scintillation timescale, scintillation bandwidth, and frequency drift rate for 12 epochs. In the remaining two cases, strong periodic modulation in the 1D frequency-domain ACF prevents reliable parameter estimation, which may be associated with AU-scale structures along the propagation path, although their locations cannot be constrained with the current data. In three observations, the secondary spectra show simultaneous detections of inner, middle, and outer arcs. By analysing the annual modulation of the inner-arc curvature, we constrain the location and velocity of the scattering screen under both isotropic and anisotropic scattering assumptions. Our results indicate that the scattering responsible for the inner arc is consistent with isotropic scattering, yielding a well-determined screen distance of $D_{\rm se, inner}=233\pm28$~pc from the Earth and its transverse velocity of $V_{\rm scr, \alpha}=-7.16\pm2.16$~km/s and $V_{\rm scr, \delta}=-41.07\pm5.69$~km/s. Analysis of the delay profiles for both the inner and outer arcs suggests that the inferred spectral exponents are consistent with, or smaller than, the Kolmogorov value. For the middle and outer arcs, under the assumption of isotropic scattering, we obtain the screen–pulsar distances of 354$\pm$22~pc and 166$\pm$12~pc, respectively.

Long-term timing analyses reveal a gradual decline in the dispersion measure (DM) of PSR~B1257+12 in recent years. From the secondary spectra, we derived the DM change rates for the three scintillation arcs. The rate for the screen closest to the pulsar is inconsistent with the observed DM decrease, whereas the rates for the middle and inner arcs are in agreement. This suggests that the DM variations are primarily dominated by plasma located farther from the pulsar. For the outer arc, the scattering screen is located roughly one-third of the way from the pulsar to Earth, yet its DM change rate is only 18\% of that of the inner arc, indicating that the region surrounding PSR~B1257+12 is relatively free of dense ionized material. Furthermore, based on the longest FAST observation of 5 hours, no scattering screens closer to the pulsar were detected, which may further imply that the pulsar’s immediate environment is relatively clean.

Our analyses suggest that the immediate environment of PSR~B1257+12 is relatively clean, with little dense ionized material in the planetary system, as indicated by the low DM change rates of the outer arc and the absence of nearby scattering screens. At the same time, detecting scattering screens closer to the pulsar may require higher-sensitivity or longer-duration observations, emphasizing the need for more precise future measurements to probe the innermost regions of the system. 

\section{Acknowledgements} 
This work is supported by the Major Science and Technology Program of Xinjiang Uygur Autonomous Region (2022A03013-2), the National Science Foundation of Xinjiang Uygur Autonomous Region (2022D01D85), the National Natural Science Foundation of China (12588202), the CAS Project for Young Scientists in Basic Research (YSBR$-$063), the Tianshan talents program (2023TSYCTD0013), the Chinese Academy of Sciences (CAS) “Light of West China” Program (No. xbzg$-$zdsys$-$202410  and No. 2022$-$XBQNXZ$-$015) and the National Key Research and Development Program of China ((No. 2022YFC2205201, No. 2022YFC2205203). The Center for Exoplanets and Habitable Worlds is supported by the Pennsylvania State University and the Eberly College of Science. D. L. acknowledges support from the New Cornerstone foundation.
\bibliography{ISS,journals,modrefs,psrrefs,crossrefs}
\bibliographystyle{aasjournal}
\end{document}